\documentclass[%
 aip,
 amsmath,amssymb,
 reprint,%
]{revtex4-2}

\usepackage{graphicx}
\usepackage{dcolumn}
\usepackage{gensymb}
\usepackage{bm}
\usepackage[dvipsnames]{xcolor}
\usepackage[pdfencoding=auto, psdextra]{hyperref}

\usepackage[utf8]{inputenc}
\usepackage[T1]{fontenc}
\usepackage{mathptmx}
\usepackage{etoolbox}

\makeatletter
\def\@email#1#2{%
 \endgroup
 \patchcmd{\titleblock@produce}
  {\frontmatter@RRAPformat}
  {\frontmatter@RRAPformat{\produce@RRAP{*#1\href{mailto:#2}{#2}}}\frontmatter@RRAPformat}
  {}{}
}%
\makeatother
\begin{document}

\preprint{AIP/123-QED}

\title[`Maser-in-a-Shoebox': a portable plug-and-play maser device at room-temperature and zero magnetic-field]{`Maser-in-a-Shoebox': a portable plug-and-play maser device at room-temperature and zero magnetic-field}
\author{Wern Ng}
   \affiliation{%
Department of Materials, Imperial College London, South Kensington SW7 2AZ, London, United Kingdom}%
  \email{wern.ng@imperial.ac.uk}
\author{Yongqiang Wen}
   \affiliation{%
Department of Materials, Imperial College London, South Kensington SW7 2AZ, London, United Kingdom}%
\author{Max Attwood}
   \affiliation{%
Department of Materials, Imperial College London, South Kensington SW7 2AZ, London, United Kingdom}%
\author{Daniel C Jones}
   \affiliation{%
Gemological Institute of America, New York, 10036, USA}%

\author{Mark Oxborrow}
   \affiliation{%
Department of Materials, Imperial College London, South Kensington SW7 2AZ, London, United Kingdom}%
\author{Neil McN. Alford}
   \affiliation{%
Department of Materials, Imperial College London, South Kensington SW7 2AZ, London, United Kingdom}%
\author{Daan M. Arroo}
   \affiliation{%
Department of Materials, Imperial College London, South Kensington SW7 2AZ, London, United Kingdom}%

\date{\today}

\begin{abstract}
Masers, the microwave analogues of lasers, have seen a renaissance owing to the discovery of gain media that mase at room-temperature and zero-applied magnetic field. However, despite the ease with which the devices can be demonstrated under ambient conditions, achieving the ubiquity and portability which lasers enjoy has to date remained challenging. We present a maser device with a miniaturized maser cavity, gain material and laser pump source that fits within the size of a shoebox. The gain medium used is pentacene-doped in para-terphenyl and it is shown to give a strong masing signal with a peak power of -5 dBm even within a smaller form factor. The device is also shown to mase at different frequencies within a small range of 1.5 MHz away from the resonant frequency. The portability and simplicity of the device, which weighs under 5 kg, paves the way for demonstrators particularly in the areas of low-noise amplifiers, quantum sensors, cavity quantum electrodynamics and long-range communications.
\end{abstract}

\maketitle

Few devices can match the exceptionally low noise temperatures of masers\cite{siegman1964microwave}, the microwave analogue of lasers. Following the discovery of the first solid-state material that could mase at room-temperature, pentacene-doped \textit{para}-terphenyl (PcPTP)\cite{Oxborrow2012}, there has been a surge in research exploring further such room-temperature maser materials and coherent microwave sources\cite{Bogatko2016,Breeze2018,gottscholl2022superradiance,attwood2023n,BimuYao2023magnonpolariton}, studying their use in ultrasensitive sensing and communication\cite{HaoWu2022enhancedQsensing,casariego2023propagating}, performance in cavity quantum electrodynamics (cQED)\cite{Breeze2017cqed,fokina2021pure,Eisenach2021,WernDAP2023}, use as polarizers for triplet dynamic nuclear polarization\cite{Kouno2019,Fujiwara2022} and as microwave mode coolers\cite{wern2021diamond,DonaldFahey2023,blank2023antimaser}. Masers have the potential to revolutionize medical imaging and long-range communications through their ultra-low noise amplification\cite{Arroo2021}, but even though room-temperature gain media have immensely simplified the maser device by removing the need for cryogenics and vacuum, the contrast between masers and their laser descendants is stark; the latter can be purchased from vendors straight to a home tabletop, while the former has remained in the domain of specialized laboratories, having not yet transcended to a more portable form. Conventional masers still weigh 100 kg and require large footprints (60 $\times$ 80 $\times$ 91 cm$^{3}$)\cite{t4sciencemaser}. Previous room-temperature maser devices in literature still required huge 100 kg electromagnets\cite{Breeze2018} or massive pump sources such as bulky xenon flash lamps\cite{HaoWu2020_cont} or pulsed optoparametric-oscillators\cite{Breeze2017cqed} which all had dimensions larger than 100 cm and weighed beyond 50 kg.

Here, we present a fully portable maser device with pumping optics and microwave cavity all contained within an enclosure the size of a shoebox (33 $\times$ 23 $\times$ 11 cm$^{3}$), weighing less than 5 kg. The gain medium used is a cylindrical single crystal (OD 3.5 mm, height 8.7 mm) of 0.1\% PcPTP grown through the Bridgman method\cite{Sloop1981,Ai2017}; PTP (Sigma-Aldrich, $\geq$99.5\%) was zone refined prior to use and mixed with pentacene at a concentration of 0.1\% (mol/mol) by grinding in a pestle and mortar. The ground powder was then loaded into a 3.5 mm OD borosilicate NMR tube and sealed under argon. Bridgman growth was performed at 4 mm/hr over 3 days at 218 $\degree$C to yield the pink PcPTP crystal.

\begin{figure*}[ht!]
\includegraphics[width=1.0\textwidth]{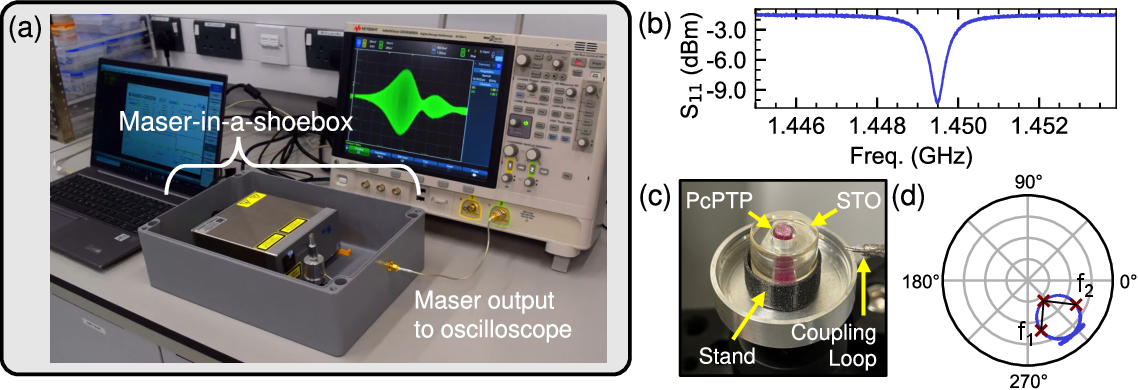}
\centering
\caption{\label{fig:shoebox}(a) Photo of the portable maser with an output masing pulse captured by direct detection on the oscilloscope. An ND-filter is seen covering the output port of the laser, but this was removed during masing experiments. (b) Example S$_{11}$ measured from the cavity coupling loop (c) Close-up photo of the inside of the aluminum cavity housing the PcPTP gain medium crystal, with various components labelled. (d) Polar plot of the S$_{11}$ plot in (b), with plot radius equal to one and showing undercoupling to the cavity.}
\end{figure*}

The critical components to miniaturize have been the microwave resonator and the optical pump for the maser. The resonator that houses the PcPTP crystal consists of an aluminum cavity enclosing a dielectric ring resonator made of a single crystal of strontium titanate (STO). The STO ring has an OD 12.2 mm, ID 4.1 mm, and height of 8.7 mm, allowing it to achieve a large Purcell factor through reducing the mode volume. This is important to reach the masing threshold. The STO supports a TE$_{01\delta}$ mode with a frequency $f_{\rm{mode}}$ of 1.4495 GHz, which could be tuned through adjusting the height of the inner ceiling of the aluminum cavity. The aluminum cavity has an ID of 22 mm and the maximum height the ceiling can be adjusted to is 20 mm. 

The laser is a diode-pumped pulsed 532 nm Nd:YAG laser (Montfort Laser GmbH, M-NANO-GREEN model PR190) with a pulse duration of 6 ns and controlled by a computer via USB. The total measured pulse energy was 30 mJ using a Thorlabs ES245C sensor, which includes both the 532 nm pulse and a residual 1064 nm laser pulse that passes along the same axis. We note that this is well above the threshold laser power required for PcPTP to mase for 532 nm light, which we estimate to be less than 7 mJ (for a 6 ns pulse) based on previously reported results\cite{WernDAP2023}. The PcPTP does not suffer any noticeable degradation due to the residual 1064 nm radiation.

Figure~\ref{fig:shoebox}(a) shows the `maser-in-a-shoebox' outputting a maser pulse into an oscilloscope. The maser cavity and pulsed laser pump source are mounted within an aluminum enclosure, which has an SMA output port installed for emitting the maser pulse into a load or measurement device of choice. Figure~\ref{fig:shoebox}(c) shows the PcPTP inserted into the STO ring (with the main body of the aluminum cavity removed to reveal the interior). The STO and PcPTP sit on a stand made of 3D-printed plastic (polylactic acid, or PLA) which supports the STO to a height of 4.5 mm above the floor of the cavity. A loop of wire on a coaxial cable acts as the coupling loop for receiving the maser pulses after the PcPTP crystal is optically pumped. The device is designed to be as `plug-and-play' as possible, where it is simply plugged into mains electricity and controlled with a computer to output maser pulses at a certain repetition rate (governed by the laser pumping repetition rate).

\begin{figure}[b]
\includegraphics[width=0.45\textwidth]{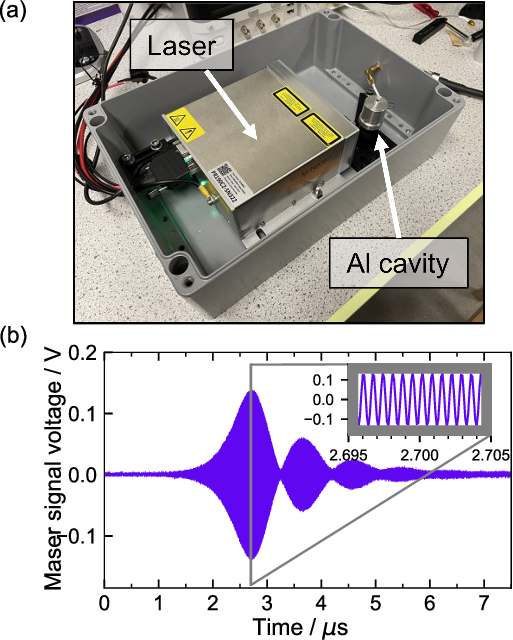}
\caption{\label{fig:mainsig}(a) View of the inside of the enclosure housing the aluminium maser cavity and laser pump. (b) Masing signal at 1.44969 GHz detected from the portable maser. The inset shows a 10 ns slice of the oscillations of the signal at the peak voltage region. The entire maser signal is also modulated by Rabi oscillations.}
\end{figure}

The laser and maser cavity are arranged as shown in Figure~\ref{fig:mainsig}(a), where the laser output is directed into the aluminum cavity through an aperture in the cavity, and passes through the STO dielectric (which has been polished to be transparent) to excite the PcPTP within. The maser operates as follows; the repetition rate of the laser is set between 0.5 to 10 Hz (or single shot). After the laser pulse hits the PcPTP, the maser burst is picked up by the coupling loop and transmitted through a coaxial cable to the output port. The $f_{\rm{mode}}$ of the resonator can be checked using the dip in the S$_{11}$ reading from a vector network analyzer (VNA) and tuned to the maser frequency $f_{\rm{res}}=1.4495$ GHz, prior to laser excitation (Figure~\ref{fig:shoebox}(b)). The loaded quality factor ($Q_{\rm{L}}$) of the resonator with coupling loop inserted can be estimated using the polar plot of the S$_{11}$ as explained in previous work\cite{WernDAP2023}. From Figure~\ref{fig:shoebox}(d), the frequencies corresponding to the -3 dB bandwidth were $f_1=1.44915$ GHz and $f_2=1.44986$ GHz respectively, giving $Q_{\rm{L}}\approx1.4495/(f_2-f_1)=2042$, with possible undercoupling as the polar plot does not intersect the center. Further explanations of how to operate masers such as PcPTP are available from previous literature\cite{WernDAP2023,Breeze2018,HaoWu2020_cont}.

Figure~\ref{fig:mainsig}(b) shows a single shot time-domain signal from the maser with the cavity tuned as close as possible to $f_{\rm{res}}$. The frequency power spectrum for this signal showed that its actual central frequency (or carrier frequency), and likely the cavity frequency, was 1.44969 GHz, which was very close to $f_{\rm{res}}$. The signal was detected directly on an oscilloscope (Keysight InfiniiVision DSOX6002A, 6 GHz bandwidth). The time at which the laser pulse triggered the oscilloscope measurement was at 0~\textmu s (detected using a Thorlabs DET10A2 photodiode). The maser pulse reaches a peak voltage of 0.13 V when measured through the oscilloscope channel impedance of 50~$\Omega$, which gives a peak power of 0.33 mW, or -5 dBm, slightly surpassing the strongest maser peak power recorded for PcPTP\cite{Breeze2017cqed}. Hence, our portable maser maintains a similar output power compared to previous implementations\cite{Oxborrow2012,Breeze2017cqed}.

Rabi oscillations are seen in the masing signal at an estimated frequency of 0.8 MHz. This is similar to what has been observed previously in literature\cite{Breeze2017cqed,WernDAP2023}, where the Rabi oscillations indicate the strong coupling regime in cQED.

\begin{figure}[h!]
\includegraphics[width=0.48\textwidth]{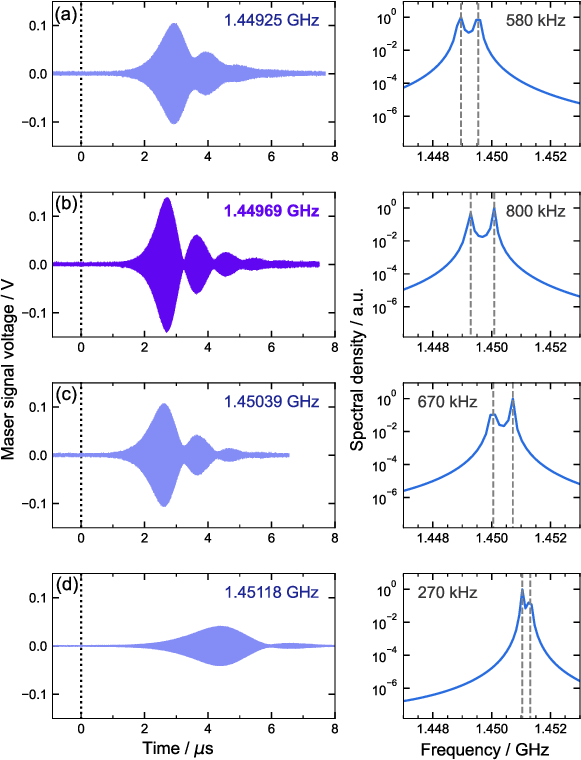}
\caption{\label{fig:time_freq}(a)-(d) Rows displaying the maser signal recorded when the cavity was tuned to different frequencies, with the grey dotted lines showing the time of the laser trigger. The right column displays the normalized frequency power spectrum of the corresponding signal in the left column, with the Rabi frequency in kHz estimated as the splitting between the two peak frequencies in each spectrum (indicated by dashed lines). The on-resonance maser signal at 1.44969 GHz (b, colored differently from the other plots) gives the greatest amplitude.}
\end{figure}

The cavity was then detuned away from $f_{\rm{res}}$ (Figure~\ref{fig:time_freq}) in order to observe the effect of detuning on the maser output power. The signal in Figure~\ref{fig:time_freq}(b) is identical to that in Figure~\ref{fig:mainsig}(b), and as expected is highest in amplitude due to 1.44969 GHz being closest to $f_{\rm{res}}$. To resolve the closely-spaced peaks associated with Rabi splitting of the maser mode around the centre frequency, power spectra were obtained from the time-domain signals using the maximum entropy method\cite{burg1975maximum}, which does not suffer from artifacts associated with arbitrary window functions used in FFT-based spectral estimation. The conversions were carried out using an implementation of Burg's method for maximum entropy spectral estimation in the \texttt{memspectrum} Python package \cite{martini2021maximum}.

Predictably, the maser power output decreases when the cavity is tuned away from $f_{\rm{res}}$. As shown in the right column of Figure~\ref{fig:time_freq}, the power spectra peaks will shift with the cavity frequency when it is tuned away from $f_{\rm{res}}$. Through monitoring the VNA, the cavity was tuned in steps of approximately 0.5 MHz and the signal was recorded. The central frequency of the power spectra of each signal corresponds to the resulting maser frequency measured.

The delay between laser excitation (at 0 \textmu s) and the first peak of the maser signal is increased the further away one is from $f_{\rm{res}}$, which is due to the stimulated transition rate decreasing when one is off-resonance. The frequency of the Rabi oscillations decreases as one moves away from $f_{\rm{res}}$, resulting in narrower splittings in the power spectra. In the case of Figure~\ref{fig:time_freq}(d), the Rabi oscillation from being off-resonance is so slow that there is only time for two cycles before spin-lattice decay has totally quenched the masing signal. This is due to the relationship between the Rabi frequency and the microwave photon number; for very large numbers of microwave photons (which is the case for masers), the Rabi frequency is proportional to the number of microwave photons in the cavity $n_{\rm{cavity}}$\cite{Eisenach2021}. As the maser is tuned away from the cavity resonance the maser amplitude decreases, and so $n_{\rm{cavity}}$ decreases as well, resulting in the Rabi frequency decreasing.

These results help to establish certain operation restrictions for the PcPTP maser; if one desires stronger coupling for cavity quantum electrodynamics, and the shortest delay to peak maser power, it is best to operate exactly on resonance where the Rabi splitting is shown to be greatest and the delay to peak power is shortest. However, if one desires a maser signal with a slightly different frequency from resonance, then this can be achieved with a small tuning window away from $f_{\rm{res}}$, corresponding to at least 1.5 MHz above or below $f_{\rm{res}}$. A maser signal would still be visible and will be at the frequency one has tuned to, but the amplitude will suffer.

In conclusion, we have demonstrated a conveniently portable maser device. Our setup also demonstrates that cavities made from metals of lower electrical conductivity, such as aluminum which is more easily machinable, are viable as maser cavities. Due to the lower conductivity of aluminum compared with copper, $Q_{\rm{L}}$ is decreased and stronger light pumping is required to compensate for this. The use of PLA plastic for the stand and tape for attaching the STO onto the stand are both lossy and decrease $Q_{\rm{L}}$; better alternatives for stands would be teflon or sapphire. However, the fact that strong masing is possible even under an unoptimized setup is a testament to the robust masing capability of our device. Our `shoebox maser' can also have its gain medium swapped for other materials, so long as they use 532 nm excitation wavelengths; the recently discovered 6,13-diazapentacene doped in para-terphenyl maser would be compatible with this wavelength while offering its faster masing startup time\cite{WernDAP2023}. We further anticipate that miniaturized room-temperature masers will find applications in quantum sensing\cite{HaoWu2022enhancedQsensing,casariego2023propagating}, quantum information processing\cite{MicrowavesinQC2021,NVcqed2011}, and serve as educational tools for demonstrating cQED phenomena just on the tabletop, all at room-temperature.

\begin{acknowledgments}
The authors dedicate this Letter to the memory of Michael Lennon, who made the cavity used for this work and whose skill in manufacturing high-Q cavities has been an important part of the story of room-temperature masers. 

We further thank Ben Gaskell of Gaskell Quartz Ltd (London) for making the STO ring used and Dr Tim Moorsom for the photographs of the maser setup. This work was supported by the U.K. Engineering and Physical Sciences Research Council through grant EP/V001914/1
\end{acknowledgments}
\section*{Author Declarations}
\subsection*{Conflict of Interest}

The authors have no conflicts to disclose.

\subsection*{Data Availability Statement}

The data that support the findings of this study are available from the corresponding author upon reasonable request.

\bibliography{aipsamp}

\begin{thebibliography}{27}%
\makeatletter
\providecommand \@ifxundefined [1]{%
 \@ifx{#1\undefined}
}%
\providecommand \@ifnum [1]{%
 \ifnum #1\expandafter \@firstoftwo
 \else \expandafter \@secondoftwo
 \fi
}%
\providecommand \@ifx [1]{%
 \ifx #1\expandafter \@firstoftwo
 \else \expandafter \@secondoftwo
 \fi
}%
\providecommand \natexlab [1]{#1}%
\providecommand \enquote  [1]{``#1''}%
\providecommand \bibnamefont  [1]{#1}%
\providecommand \bibfnamefont [1]{#1}%
\providecommand \citenamefont [1]{#1}%
\providecommand \href@noop [0]{\@secondoftwo}%
\providecommand \href [0]{\begingroup \@sanitize@url \@href}%
\providecommand \@href[1]{\@@startlink{#1}\@@href}%
\providecommand \@@href[1]{\endgroup#1\@@endlink}%
\providecommand \@sanitize@url [0]{\catcode `\\12\catcode `\$12\catcode
  `\&12\catcode `\#12\catcode `\^12\catcode `\_12\catcode `\%12\relax}%
\providecommand \@@startlink[1]{}%
\providecommand \@@endlink[0]{}%
\providecommand \url  [0]{\begingroup\@sanitize@url \@url }%
\providecommand \@url [1]{\endgroup\@href {#1}{\urlprefix }}%
\providecommand \urlprefix  [0]{URL }%
\providecommand \Eprint [0]{\href }%
\providecommand \doibase [0]{https://doi.org/}%
\providecommand \selectlanguage [0]{\@gobble}%
\providecommand \bibinfo  [0]{\@secondoftwo}%
\providecommand \bibfield  [0]{\@secondoftwo}%
\providecommand \translation [1]{[#1]}%
\providecommand \BibitemOpen [0]{}%
\providecommand \bibitemStop [0]{}%
\providecommand \bibitemNoStop [0]{.\EOS\space}%
\providecommand \EOS [0]{\spacefactor3000\relax}%
\providecommand \BibitemShut  [1]{\csname bibitem#1\endcsname}%
\let\auto@bib@innerbib\@empty
\bibitem [{\citenamefont {Siegmann}(1964)}]{siegman1964microwave}%
  \BibitemOpen
  \bibfield  {author} {\bibinfo {author} {\bibfnamefont {A.~E.}\ \bibnamefont
  {Siegmann}},\ }\href@noop {} {\emph {\bibinfo {title} {Microwave Solid-State
  Masers}}}\ (\bibinfo  {publisher} {McGraw-Hill},\ \bibinfo {year}
  {1964})\BibitemShut {NoStop}%
\bibitem [{\citenamefont {Oxborrow}, \citenamefont {Breeze},\ and\
  \citenamefont {Alford}(2012)}]{Oxborrow2012}%
  \BibitemOpen
  \bibfield  {author} {\bibinfo {author} {\bibfnamefont {M.}~\bibnamefont
  {Oxborrow}}, \bibinfo {author} {\bibfnamefont {J.~D.}\ \bibnamefont
  {Breeze}},\ and\ \bibinfo {author} {\bibfnamefont {N.~M.}\ \bibnamefont
  {Alford}},\ }\bibfield  {title} {\enquote {\bibinfo {title}
  {{Room-temperature solid-state maser}},}\ }\href
  {https://doi.org/10.1038/nature11339} {\bibfield  {journal} {\bibinfo
  {journal} {Nature}\ }\textbf {\bibinfo {volume} {488}},\ \bibinfo {pages}
  {353--356} (\bibinfo {year} {2012})}\BibitemShut {NoStop}%
\bibitem [{\citenamefont {Bogatko}\ \emph {et~al.}(2016)\citenamefont
  {Bogatko}, \citenamefont {Haynes}, \citenamefont {Sathian}, \citenamefont
  {Wade}, \citenamefont {Kim}, \citenamefont {Tan}, \citenamefont {Breeze},
  \citenamefont {Salvadori}, \citenamefont {Horsfield},\ and\ \citenamefont
  {Oxborrow}}]{Bogatko2016}%
  \BibitemOpen
  \bibfield  {author} {\bibinfo {author} {\bibfnamefont {S.}~\bibnamefont
  {Bogatko}}, \bibinfo {author} {\bibfnamefont {P.~D.}\ \bibnamefont {Haynes}},
  \bibinfo {author} {\bibfnamefont {J.}~\bibnamefont {Sathian}}, \bibinfo
  {author} {\bibfnamefont {J.}~\bibnamefont {Wade}}, \bibinfo {author}
  {\bibfnamefont {J.~S.}\ \bibnamefont {Kim}}, \bibinfo {author} {\bibfnamefont
  {K.~J.}\ \bibnamefont {Tan}}, \bibinfo {author} {\bibfnamefont
  {J.}~\bibnamefont {Breeze}}, \bibinfo {author} {\bibfnamefont
  {E.}~\bibnamefont {Salvadori}}, \bibinfo {author} {\bibfnamefont
  {A.}~\bibnamefont {Horsfield}},\ and\ \bibinfo {author} {\bibfnamefont
  {M.}~\bibnamefont {Oxborrow}},\ }\bibfield  {title} {\enquote {\bibinfo
  {title} {{Molecular Design of a Room-Temperature Maser}},}\ }\href
  {https://doi.org/10.1021/acs.jpcc.6b00150} {\bibfield  {journal} {\bibinfo
  {journal} {J. Phys. Chem. C}\ }\textbf {\bibinfo {volume} {120}},\ \bibinfo
  {pages} {8251--8260} (\bibinfo {year} {2016})}\BibitemShut {NoStop}%
\bibitem [{\citenamefont {Breeze}\ \emph {et~al.}(2018)\citenamefont {Breeze},
  \citenamefont {Salvadori}, \citenamefont {Sathian}, \citenamefont {Alford},\
  and\ \citenamefont {Kay}}]{Breeze2018}%
  \BibitemOpen
  \bibfield  {author} {\bibinfo {author} {\bibfnamefont {J.}~\bibnamefont
  {Breeze}}, \bibinfo {author} {\bibfnamefont {E.}~\bibnamefont {Salvadori}},
  \bibinfo {author} {\bibfnamefont {J.}~\bibnamefont {Sathian}}, \bibinfo
  {author} {\bibfnamefont {N.~M.}\ \bibnamefont {Alford}},\ and\ \bibinfo
  {author} {\bibfnamefont {C.~W.~M.}\ \bibnamefont {Kay}},\ }\bibfield  {title}
  {\enquote {\bibinfo {title} {Continuous-wave room-temperature diamond
  maser},}\ }\href@noop {} {\bibfield  {journal} {\bibinfo  {journal} {Nature}\
  }\textbf {\bibinfo {volume} {555}},\ \bibinfo {pages} {493--496} (\bibinfo
  {year} {2018})}\BibitemShut {NoStop}%
\bibitem [{\citenamefont {Gottscholl}\ \emph {et~al.}(2022)\citenamefont
  {Gottscholl}, \citenamefont {Wagenh{\"o}fer}, \citenamefont {Klimmer},
  \citenamefont {Scherbel}, \citenamefont {Kasper}, \citenamefont {Baianov},
  \citenamefont {Astakhov}, \citenamefont {Dyakonov},\ and\ \citenamefont
  {Sperlich}}]{gottscholl2022superradiance}%
  \BibitemOpen
  \bibfield  {author} {\bibinfo {author} {\bibfnamefont {A.}~\bibnamefont
  {Gottscholl}}, \bibinfo {author} {\bibfnamefont {M.}~\bibnamefont
  {Wagenh{\"o}fer}}, \bibinfo {author} {\bibfnamefont {M.}~\bibnamefont
  {Klimmer}}, \bibinfo {author} {\bibfnamefont {S.}~\bibnamefont {Scherbel}},
  \bibinfo {author} {\bibfnamefont {C.}~\bibnamefont {Kasper}}, \bibinfo
  {author} {\bibfnamefont {V.}~\bibnamefont {Baianov}}, \bibinfo {author}
  {\bibfnamefont {G.~V.}\ \bibnamefont {Astakhov}}, \bibinfo {author}
  {\bibfnamefont {V.}~\bibnamefont {Dyakonov}},\ and\ \bibinfo {author}
  {\bibfnamefont {A.}~\bibnamefont {Sperlich}},\ }\bibfield  {title} {\enquote
  {\bibinfo {title} {Superradiance of spin defects in silicon carbide for maser
  applications},}\ }\href {https://doi.org/10.3389/fphot.2022.886354}
  {\bibfield  {journal} {\bibinfo  {journal} {Front. Photon.}\ }\textbf
  {\bibinfo {volume} {3}},\ \bibinfo {pages} {886354} (\bibinfo {year}
  {2022})}\BibitemShut {NoStop}%
\bibitem [{\citenamefont {Attwood}\ \emph {et~al.}(2023)\citenamefont
  {Attwood}, \citenamefont {Xu}, \citenamefont {Newns}, \citenamefont {Meng},
  \citenamefont {Ingle}, \citenamefont {Wu}, \citenamefont {Chen},
  \citenamefont {Xu}, \citenamefont {Ng}, \citenamefont {Abiola}, \citenamefont
  {Stavros},\ and\ \citenamefont {Oxborrow}}]{attwood2023n}%
  \BibitemOpen
  \bibfield  {author} {\bibinfo {author} {\bibfnamefont {M.}~\bibnamefont
  {Attwood}}, \bibinfo {author} {\bibfnamefont {X.}~\bibnamefont {Xu}},
  \bibinfo {author} {\bibfnamefont {M.}~\bibnamefont {Newns}}, \bibinfo
  {author} {\bibfnamefont {Z.}~\bibnamefont {Meng}}, \bibinfo {author}
  {\bibfnamefont {R.~A.}\ \bibnamefont {Ingle}}, \bibinfo {author}
  {\bibfnamefont {H.}~\bibnamefont {Wu}}, \bibinfo {author} {\bibfnamefont
  {X.}~\bibnamefont {Chen}}, \bibinfo {author} {\bibfnamefont {W.}~\bibnamefont
  {Xu}}, \bibinfo {author} {\bibfnamefont {W.}~\bibnamefont {Ng}}, \bibinfo
  {author} {\bibfnamefont {T.~T.}\ \bibnamefont {Abiola}}, \bibinfo {author}
  {\bibfnamefont {V.~G.}\ \bibnamefont {Stavros}},\ and\ \bibinfo {author}
  {\bibfnamefont {M.}~\bibnamefont {Oxborrow}},\ }\bibfield  {title} {\enquote
  {\bibinfo {title} {N-heteroacenes as an organic gain medium for
  room-temperature masers},}\ }\href
  {https://doi.org/10.1021/acs.chemmater.3c00640} {\bibfield  {journal}
  {\bibinfo  {journal} {Chem. Mater.}\ }\textbf {\bibinfo {volume} {35}},\
  \bibinfo {pages} {4498--4509} (\bibinfo {year} {2023})},\ \Eprint
  {https://arxiv.org/abs/https://doi.org/10.1021/acs.chemmater.3c00640}
  {https://doi.org/10.1021/acs.chemmater.3c00640} \BibitemShut {NoStop}%
\bibitem [{\citenamefont {Yao}\ \emph {et~al.}(2023)\citenamefont {Yao},
  \citenamefont {Gui}, \citenamefont {Rao}, \citenamefont {Zhang},
  \citenamefont {Lu},\ and\ \citenamefont {Hu}}]{BimuYao2023magnonpolariton}%
  \BibitemOpen
  \bibfield  {author} {\bibinfo {author} {\bibfnamefont {B.}~\bibnamefont
  {Yao}}, \bibinfo {author} {\bibfnamefont {Y.~S.}\ \bibnamefont {Gui}},
  \bibinfo {author} {\bibfnamefont {J.~W.}\ \bibnamefont {Rao}}, \bibinfo
  {author} {\bibfnamefont {Y.~H.}\ \bibnamefont {Zhang}}, \bibinfo {author}
  {\bibfnamefont {W.}~\bibnamefont {Lu}},\ and\ \bibinfo {author}
  {\bibfnamefont {C.-M.}\ \bibnamefont {Hu}},\ }\bibfield  {title} {\enquote
  {\bibinfo {title} {Coherent microwave emission of gain-driven polaritons},}\
  }\href {https://doi.org/10.1103/PhysRevLett.130.146702} {\bibfield  {journal}
  {\bibinfo  {journal} {Phys. Rev. Lett.}\ }\textbf {\bibinfo {volume} {130}},\
  \bibinfo {pages} {146702} (\bibinfo {year} {2023})}\BibitemShut {NoStop}%
\bibitem [{\citenamefont {Wu}\ \emph {et~al.}(2022)\citenamefont {Wu},
  \citenamefont {Yang}, \citenamefont {Oxborrow}, \citenamefont {Jiang},
  \citenamefont {Zhao}, \citenamefont {Budker}, \citenamefont {Zhang},\ and\
  \citenamefont {Du}}]{HaoWu2022enhancedQsensing}%
  \BibitemOpen
  \bibfield  {author} {\bibinfo {author} {\bibfnamefont {H.}~\bibnamefont
  {Wu}}, \bibinfo {author} {\bibfnamefont {S.}~\bibnamefont {Yang}}, \bibinfo
  {author} {\bibfnamefont {M.}~\bibnamefont {Oxborrow}}, \bibinfo {author}
  {\bibfnamefont {M.}~\bibnamefont {Jiang}}, \bibinfo {author} {\bibfnamefont
  {Q.}~\bibnamefont {Zhao}}, \bibinfo {author} {\bibfnamefont {D.}~\bibnamefont
  {Budker}}, \bibinfo {author} {\bibfnamefont {B.}~\bibnamefont {Zhang}},\ and\
  \bibinfo {author} {\bibfnamefont {J.}~\bibnamefont {Du}},\ }\bibfield
  {title} {\enquote {\bibinfo {title} {Enhanced quantum sensing with
  room-temperature solid-state masers},}\ }\href
  {https://doi.org/10.1126/sciadv.ade1613} {\bibfield  {journal} {\bibinfo
  {journal} {Sci. Adv.}\ }\textbf {\bibinfo {volume} {8}},\ \bibinfo {pages}
  {eade1613} (\bibinfo {year} {2022})}\BibitemShut {NoStop}%
\bibitem [{\citenamefont {Casariego}\ \emph {et~al.}(2023)\citenamefont
  {Casariego}, \citenamefont {Cruzeiro}, \citenamefont {Gherardini},
  \citenamefont {Gonzalez-Raya}, \citenamefont {Andr{\'e}}, \citenamefont
  {Fraz{\~a}o}, \citenamefont {Catto}, \citenamefont {M{\"o}tt{\"o}nen},
  \citenamefont {Datta}, \citenamefont {Viisanen} \emph
  {et~al.}}]{casariego2023propagating}%
  \BibitemOpen
  \bibfield  {author} {\bibinfo {author} {\bibfnamefont {M.}~\bibnamefont
  {Casariego}}, \bibinfo {author} {\bibfnamefont {E.~Z.}\ \bibnamefont
  {Cruzeiro}}, \bibinfo {author} {\bibfnamefont {S.}~\bibnamefont
  {Gherardini}}, \bibinfo {author} {\bibfnamefont {T.}~\bibnamefont
  {Gonzalez-Raya}}, \bibinfo {author} {\bibfnamefont {R.}~\bibnamefont
  {Andr{\'e}}}, \bibinfo {author} {\bibfnamefont {G.}~\bibnamefont
  {Fraz{\~a}o}}, \bibinfo {author} {\bibfnamefont {G.}~\bibnamefont {Catto}},
  \bibinfo {author} {\bibfnamefont {M.}~\bibnamefont {M{\"o}tt{\"o}nen}},
  \bibinfo {author} {\bibfnamefont {D.}~\bibnamefont {Datta}}, \bibinfo
  {author} {\bibfnamefont {K.}~\bibnamefont {Viisanen}}, \emph {et~al.},\
  }\bibfield  {title} {\enquote {\bibinfo {title} {Propagating quantum
  microwaves: towards applications in communication and sensing},}\ }\href
  {https://doi.org/10.1088/2058-9565/acc4af} {\bibfield  {journal} {\bibinfo
  {journal} {Quantum Science and Technology}\ }\textbf {\bibinfo {volume}
  {8}},\ \bibinfo {pages} {023001} (\bibinfo {year} {2023})}\BibitemShut
  {NoStop}%
\bibitem [{\citenamefont {Breeze}\ \emph {et~al.}(2017)\citenamefont {Breeze},
  \citenamefont {Salvadori}, \citenamefont {Sathian}, \citenamefont {Alford},\
  and\ \citenamefont {Kay}}]{Breeze2017cqed}%
  \BibitemOpen
  \bibfield  {author} {\bibinfo {author} {\bibfnamefont {J.~D.}\ \bibnamefont
  {Breeze}}, \bibinfo {author} {\bibfnamefont {E.}~\bibnamefont {Salvadori}},
  \bibinfo {author} {\bibfnamefont {J.}~\bibnamefont {Sathian}}, \bibinfo
  {author} {\bibfnamefont {N.~M.}\ \bibnamefont {Alford}},\ and\ \bibinfo
  {author} {\bibfnamefont {C.~W.~M.}\ \bibnamefont {Kay}},\ }\bibfield  {title}
  {\enquote {\bibinfo {title} {Room-temperature cavity quantum electrodynamics
  with strongly coupled {D}icke states},}\ }\href
  {https://doi.org/10.1038/s41534-017-0041-3} {\bibfield  {journal} {\bibinfo
  {journal} {npj Quantum Inf.}\ }\textbf {\bibinfo {volume} {3}},\ \bibinfo
  {pages} {40} (\bibinfo {year} {2017})}\BibitemShut {NoStop}%
\bibitem [{\citenamefont {Fokina}\ and\ \citenamefont
  {Elizbarashvili}(2021)}]{fokina2021pure}%
  \BibitemOpen
  \bibfield  {author} {\bibinfo {author} {\bibfnamefont {N.}~\bibnamefont
  {Fokina}}\ and\ \bibinfo {author} {\bibfnamefont {M.}~\bibnamefont
  {Elizbarashvili}},\ }\bibfield  {title} {\enquote {\bibinfo {title} {Pure
  superradiance from the inverted levels of spin triplet states coupled to
  resonator},}\ }\href {https://doi.org/10.1007/s00723-021-01346-x} {\bibfield
  {journal} {\bibinfo  {journal} {Appl. Magn. Reson.}\ }\textbf {\bibinfo
  {volume} {52}},\ \bibinfo {pages} {769--780} (\bibinfo {year}
  {2021})}\BibitemShut {NoStop}%
\bibitem [{\citenamefont {Eisenach}\ \emph {et~al.}(2021)\citenamefont
  {Eisenach}, \citenamefont {Barry}, \citenamefont {O'Keeffe}, \citenamefont
  {Schloss}, \citenamefont {Steinecker}, \citenamefont {Englund},\ and\
  \citenamefont {Braje}}]{Eisenach2021}%
  \BibitemOpen
  \bibfield  {author} {\bibinfo {author} {\bibfnamefont {E.~R.}\ \bibnamefont
  {Eisenach}}, \bibinfo {author} {\bibfnamefont {J.~F.}\ \bibnamefont {Barry}},
  \bibinfo {author} {\bibfnamefont {M.~F.}\ \bibnamefont {O'Keeffe}}, \bibinfo
  {author} {\bibfnamefont {J.~M.}\ \bibnamefont {Schloss}}, \bibinfo {author}
  {\bibfnamefont {M.~H.}\ \bibnamefont {Steinecker}}, \bibinfo {author}
  {\bibfnamefont {D.~R.}\ \bibnamefont {Englund}},\ and\ \bibinfo {author}
  {\bibfnamefont {D.~A.}\ \bibnamefont {Braje}},\ }\bibfield  {title} {\enquote
  {\bibinfo {title} {Cavity-enhanced microwave readout of a solid-state spin
  sensor},}\ }\href {https://doi.org/10.1038/s41467-021-21256-7} {\bibfield
  {journal} {\bibinfo  {journal} {Nat. Commun.}\ }\textbf {\bibinfo {volume}
  {12}},\ \bibinfo {pages} {1357} (\bibinfo {year} {2021})}\BibitemShut
  {NoStop}%
\bibitem [{\citenamefont {Ng}\ \emph {et~al.}(2023)\citenamefont {Ng},
  \citenamefont {Xu}, \citenamefont {Attwood}, \citenamefont {Wu},
  \citenamefont {Meng}, \citenamefont {Chen},\ and\ \citenamefont
  {Oxborrow}}]{WernDAP2023}%
  \BibitemOpen
  \bibfield  {author} {\bibinfo {author} {\bibfnamefont {W.}~\bibnamefont
  {Ng}}, \bibinfo {author} {\bibfnamefont {X.}~\bibnamefont {Xu}}, \bibinfo
  {author} {\bibfnamefont {M.}~\bibnamefont {Attwood}}, \bibinfo {author}
  {\bibfnamefont {H.}~\bibnamefont {Wu}}, \bibinfo {author} {\bibfnamefont
  {Z.}~\bibnamefont {Meng}}, \bibinfo {author} {\bibfnamefont {X.}~\bibnamefont
  {Chen}},\ and\ \bibinfo {author} {\bibfnamefont {M.}~\bibnamefont
  {Oxborrow}},\ }\bibfield  {title} {\enquote {\bibinfo {title} {Move aside
  pentacene: Diazapentacene-doped para-terphenyl, a zero-field room-temperature
  maser with strong coupling for cavity quantum electrodynamics},}\ }\href
  {https://doi.org/https://doi.org/10.1002/adma.202300441} {\bibfield
  {journal} {\bibinfo  {journal} {Advanced Materials}\ }\textbf {\bibinfo
  {volume} {35}},\ \bibinfo {pages} {2300441} (\bibinfo {year} {2023})},\
  \Eprint
  {https://arxiv.org/abs/https://onlinelibrary.wiley.com/doi/pdf/10.1002/adma.202300441}
  {https://onlinelibrary.wiley.com/doi/pdf/10.1002/adma.202300441} \BibitemShut
  {NoStop}%
\bibitem [{\citenamefont {Kouno}\ \emph {et~al.}(2019)\citenamefont {Kouno},
  \citenamefont {Kawashima}, \citenamefont {Tateishi}, \citenamefont {Uesaka},
  \citenamefont {Kimizuka},\ and\ \citenamefont {Yanai}}]{Kouno2019}%
  \BibitemOpen
  \bibfield  {author} {\bibinfo {author} {\bibfnamefont {H.}~\bibnamefont
  {Kouno}}, \bibinfo {author} {\bibfnamefont {Y.}~\bibnamefont {Kawashima}},
  \bibinfo {author} {\bibfnamefont {K.}~\bibnamefont {Tateishi}}, \bibinfo
  {author} {\bibfnamefont {T.}~\bibnamefont {Uesaka}}, \bibinfo {author}
  {\bibfnamefont {N.}~\bibnamefont {Kimizuka}},\ and\ \bibinfo {author}
  {\bibfnamefont {N.}~\bibnamefont {Yanai}},\ }\bibfield  {title} {\enquote
  {\bibinfo {title} {Nonpentacene polarizing agents with improved air stability
  for triplet dynamic nuclear polarization at room temperature},}\ }\href@noop
  {} {\bibfield  {journal} {\bibinfo  {journal} {J. Phys. Chem. Lett.}\
  }\textbf {\bibinfo {volume} {10}},\ \bibinfo {pages} {2208--2213} (\bibinfo
  {year} {2019})}\BibitemShut {NoStop}%
\bibitem [{\citenamefont {Fujiwara}\ \emph {et~al.}(2022)\citenamefont
  {Fujiwara}, \citenamefont {Matsumoto}, \citenamefont {Nishimura},
  \citenamefont {Kimizuka}, \citenamefont {Tateishi}, \citenamefont {Uesaka},\
  and\ \citenamefont {Yanai}}]{Fujiwara2022}%
  \BibitemOpen
  \bibfield  {author} {\bibinfo {author} {\bibfnamefont {S.}~\bibnamefont
  {Fujiwara}}, \bibinfo {author} {\bibfnamefont {N.}~\bibnamefont {Matsumoto}},
  \bibinfo {author} {\bibfnamefont {K.}~\bibnamefont {Nishimura}}, \bibinfo
  {author} {\bibfnamefont {N.}~\bibnamefont {Kimizuka}}, \bibinfo {author}
  {\bibfnamefont {K.}~\bibnamefont {Tateishi}}, \bibinfo {author}
  {\bibfnamefont {T.}~\bibnamefont {Uesaka}},\ and\ \bibinfo {author}
  {\bibfnamefont {N.}~\bibnamefont {Yanai}},\ }\bibfield  {title} {\enquote
  {\bibinfo {title} {Triplet dynamic nuclear polarization of guest molecules
  through induced fit in a flexible metal–organic framework},}\ }\href
  {https://doi.org/https://doi.org/10.1002/anie.202115792} {\bibfield
  {journal} {\bibinfo  {journal} {Angew. Chem. Int. Ed.}\ }\textbf {\bibinfo
  {volume} {61}},\ \bibinfo {pages} {e202115792} (\bibinfo {year} {2022})},\
  \Eprint
  {https://arxiv.org/abs/https://onlinelibrary.wiley.com/doi/pdf/10.1002/anie.202115792}
  {https://onlinelibrary.wiley.com/doi/pdf/10.1002/anie.202115792} \BibitemShut
  {NoStop}%
\bibitem [{\citenamefont {Ng}, \citenamefont {Wu},\ and\ \citenamefont
  {Oxborrow}(2021)}]{wern2021diamond}%
  \BibitemOpen
  \bibfield  {author} {\bibinfo {author} {\bibfnamefont {W.}~\bibnamefont
  {Ng}}, \bibinfo {author} {\bibfnamefont {H.}~\bibnamefont {Wu}},\ and\
  \bibinfo {author} {\bibfnamefont {M.}~\bibnamefont {Oxborrow}},\ }\bibfield
  {title} {\enquote {\bibinfo {title} {{Quasi-continuous cooling of a microwave
  mode on a benchtop using hyperpolarized NV$^-$ diamond}},}\ }\href
  {https://doi.org/10.1063/5.0076460} {\bibfield  {journal} {\bibinfo
  {journal} {Appl. Phys. Lett.}\ }\textbf {\bibinfo {volume} {119}},\ \bibinfo
  {pages} {234001} (\bibinfo {year} {2021})}\BibitemShut {NoStop}%
\bibitem [{\citenamefont {Fahey}\ \emph {et~al.}(2023)\citenamefont {Fahey},
  \citenamefont {Jacobs}, \citenamefont {Turner}, \citenamefont {Choi},
  \citenamefont {Hoffman}, \citenamefont {Englund},\ and\ \citenamefont
  {Trusheim}}]{DonaldFahey2023}%
  \BibitemOpen
  \bibfield  {author} {\bibinfo {author} {\bibfnamefont {D.~P.}\ \bibnamefont
  {Fahey}}, \bibinfo {author} {\bibfnamefont {K.}~\bibnamefont {Jacobs}},
  \bibinfo {author} {\bibfnamefont {M.~J.}\ \bibnamefont {Turner}}, \bibinfo
  {author} {\bibfnamefont {H.}~\bibnamefont {Choi}}, \bibinfo {author}
  {\bibfnamefont {J.~E.}\ \bibnamefont {Hoffman}}, \bibinfo {author}
  {\bibfnamefont {D.}~\bibnamefont {Englund}},\ and\ \bibinfo {author}
  {\bibfnamefont {M.~E.}\ \bibnamefont {Trusheim}},\ }\bibfield  {title}
  {\enquote {\bibinfo {title} {Steady-state microwave mode cooling with a
  diamond {N-$V$} ensemble},}\ }\href
  {https://doi.org/10.1103/PhysRevApplied.20.014033} {\bibfield  {journal}
  {\bibinfo  {journal} {Phys. Rev. Appl.}\ }\textbf {\bibinfo {volume} {20}},\
  \bibinfo {pages} {014033} (\bibinfo {year} {2023})}\BibitemShut {NoStop}%
\bibitem [{\citenamefont {Blank}\ \emph {et~al.}(2023)\citenamefont {Blank},
  \citenamefont {Sherman}, \citenamefont {Koren},\ and\ \citenamefont
  {Zgadzai}}]{blank2023antimaser}%
  \BibitemOpen
  \bibfield  {author} {\bibinfo {author} {\bibfnamefont {A.}~\bibnamefont
  {Blank}}, \bibinfo {author} {\bibfnamefont {A.}~\bibnamefont {Sherman}},
  \bibinfo {author} {\bibfnamefont {B.}~\bibnamefont {Koren}},\ and\ \bibinfo
  {author} {\bibfnamefont {O.}~\bibnamefont {Zgadzai}},\ }\href@noop {}
  {\enquote {\bibinfo {title} {An anti-maser for quantum-limited cooling of a
  microwave cavity},}\ } (\bibinfo {year} {2023}),\ \Eprint
  {https://arxiv.org/abs/2307.12691} {arXiv:2307.12691 [quant-ph]} \BibitemShut
  {NoStop}%
\bibitem [{\citenamefont {Arroo}, \citenamefont {Alford},\ and\ \citenamefont
  {Breeze}(2021)}]{Arroo2021}%
  \BibitemOpen
  \bibfield  {author} {\bibinfo {author} {\bibfnamefont {D.~M.}\ \bibnamefont
  {Arroo}}, \bibinfo {author} {\bibfnamefont {N.~M.}\ \bibnamefont {Alford}},\
  and\ \bibinfo {author} {\bibfnamefont {J.~D.}\ \bibnamefont {Breeze}},\
  }\bibfield  {title} {\enquote {\bibinfo {title} {{Perspective on
  room-temperature solid-state masers}},}\ }\href
  {https://doi.org/10.1063/5.0061330} {\bibfield  {journal} {\bibinfo
  {journal} {Appl. Phys. Lett.}\ }\textbf {\bibinfo {volume} {119}} (\bibinfo
  {year} {2021}),\ 10.1063/5.0061330},\ \bibinfo {note} {140502}\BibitemShut
  {NoStop}%
\bibitem [{\citenamefont {T4Science}()}]{t4sciencemaser}%
  \BibitemOpen
  \bibfield  {author} {\bibinfo {author} {\bibnamefont {T4Science}},\
  }\href@noop {} {\enquote {\bibinfo {title} {i{M}aser 3000},}\ }\bibinfo
  {howpublished} {https://www.t4science.ch/products/imaser3000/},\ \bibinfo
  {note} {(Accessed: 2023-10-10)}\BibitemShut {NoStop}%
\bibitem [{\citenamefont {Wu}\ \emph {et~al.}(2020)\citenamefont {Wu},
  \citenamefont {Xie}, \citenamefont {Ng}, \citenamefont {Mehanna},
  \citenamefont {Li}, \citenamefont {Attwood},\ and\ \citenamefont
  {Oxborrow}}]{HaoWu2020_cont}%
  \BibitemOpen
  \bibfield  {author} {\bibinfo {author} {\bibfnamefont {H.}~\bibnamefont
  {Wu}}, \bibinfo {author} {\bibfnamefont {X.}~\bibnamefont {Xie}}, \bibinfo
  {author} {\bibfnamefont {W.}~\bibnamefont {Ng}}, \bibinfo {author}
  {\bibfnamefont {S.}~\bibnamefont {Mehanna}}, \bibinfo {author} {\bibfnamefont
  {Y.}~\bibnamefont {Li}}, \bibinfo {author} {\bibfnamefont {M.}~\bibnamefont
  {Attwood}},\ and\ \bibinfo {author} {\bibfnamefont {M.}~\bibnamefont
  {Oxborrow}},\ }\bibfield  {title} {\enquote {\bibinfo {title}
  {Room-temperature quasi-continuous-wave pentacene maser pumped by an invasive
  $\mathrm{Ce}:\mathrm{YAG}$ luminescent concentrator},}\ }\href
  {https://doi.org/10.1103/PhysRevApplied.14.064017} {\bibfield  {journal}
  {\bibinfo  {journal} {Phys. Rev. Applied}\ }\textbf {\bibinfo {volume}
  {14}},\ \bibinfo {pages} {064017} (\bibinfo {year} {2020})}\BibitemShut
  {NoStop}%
\bibitem [{\citenamefont {Sloop}\ \emph {et~al.}(1981)\citenamefont {Sloop},
  \citenamefont {Yu}, \citenamefont {Lin},\ and\ \citenamefont
  {Weissman}}]{Sloop1981}%
  \BibitemOpen
  \bibfield  {author} {\bibinfo {author} {\bibfnamefont {D.~J.}\ \bibnamefont
  {Sloop}}, \bibinfo {author} {\bibfnamefont {H.}~\bibnamefont {Yu}}, \bibinfo
  {author} {\bibfnamefont {T.}~\bibnamefont {Lin}},\ and\ \bibinfo {author}
  {\bibfnamefont {S.~I.}\ \bibnamefont {Weissman}},\ }\bibfield  {title}
  {\enquote {\bibinfo {title} {Electron spin echoes of a photoexcited triplet:
  Pentacene in p‐terphenyl crystals},}\ }\href
  {https://doi.org/10.1063/1.442520} {\bibfield  {journal} {\bibinfo  {journal}
  {J. Chem. Phys.}\ }\textbf {\bibinfo {volume} {75}},\ \bibinfo {pages}
  {3746--3757} (\bibinfo {year} {1981})}\BibitemShut {NoStop}%
\bibitem [{\citenamefont {Ai}\ \emph {et~al.}(2017)\citenamefont {Ai},
  \citenamefont {Chen}, \citenamefont {Feng},\ and\ \citenamefont
  {Xu}}]{Ai2017}%
  \BibitemOpen
  \bibfield  {author} {\bibinfo {author} {\bibfnamefont {Q.}~\bibnamefont
  {Ai}}, \bibinfo {author} {\bibfnamefont {P.}~\bibnamefont {Chen}}, \bibinfo
  {author} {\bibfnamefont {Y.}~\bibnamefont {Feng}},\ and\ \bibinfo {author}
  {\bibfnamefont {Y.}~\bibnamefont {Xu}},\ }\bibfield  {title} {\enquote
  {\bibinfo {title} {Growth of pentacene-doped p-terphenyl crystals by vertical
  bridgman technique and doping effect on their characterization},}\ }\href
  {https://doi.org/10.1021/acs.cgd.6b01900} {\bibfield  {journal} {\bibinfo
  {journal} {Cryst. Growth Des.}\ }\textbf {\bibinfo {volume} {17}},\ \bibinfo
  {pages} {2473--2477} (\bibinfo {year} {2017})}\BibitemShut {NoStop}%
\bibitem [{\citenamefont {Burg}(1975)}]{burg1975maximum}%
  \BibitemOpen
  \bibfield  {author} {\bibinfo {author} {\bibfnamefont {J.~P.}\ \bibnamefont
  {Burg}},\ }\href
  {https://sepwww.stanford.edu/data/media/public/oldreports/sep06/} {\emph
  {\bibinfo {title} {Maximum entropy spectral analysis.}}}\ (\bibinfo
  {publisher} {Stanford University},\ \bibinfo {year} {1975})\BibitemShut
  {NoStop}%
\bibitem [{\citenamefont {Martini}, \citenamefont {Schmidt},\ and\
  \citenamefont {Del~Pozzo}(2021)}]{martini2021maximum}%
  \BibitemOpen
  \bibfield  {author} {\bibinfo {author} {\bibfnamefont {A.}~\bibnamefont
  {Martini}}, \bibinfo {author} {\bibfnamefont {S.}~\bibnamefont {Schmidt}},\
  and\ \bibinfo {author} {\bibfnamefont {W.}~\bibnamefont {Del~Pozzo}},\
  }\bibfield  {title} {\enquote {\bibinfo {title} {Maximum entropy spectral
  analysis: A case study},}\ }\href {https://arxiv.org/abs/2106.09499}
  {\bibfield  {journal} {\bibinfo  {journal} {arXiv:2106.09499}\ } (\bibinfo
  {year} {2021})}\BibitemShut {NoStop}%
\bibitem [{\citenamefont {Bardin}, \citenamefont {Slichter},\ and\
  \citenamefont {Reilly}(2021)}]{MicrowavesinQC2021}%
  \BibitemOpen
  \bibfield  {author} {\bibinfo {author} {\bibfnamefont {J.~C.}\ \bibnamefont
  {Bardin}}, \bibinfo {author} {\bibfnamefont {D.~H.}\ \bibnamefont
  {Slichter}},\ and\ \bibinfo {author} {\bibfnamefont {D.~J.}\ \bibnamefont
  {Reilly}},\ }\bibfield  {title} {\enquote {\bibinfo {title} {Microwaves in
  quantum computing},}\ }\href {https://doi.org/10.1109/JMW.2020.3034071}
  {\bibfield  {journal} {\bibinfo  {journal} {IEEE Journal of Microwaves}\
  }\textbf {\bibinfo {volume} {1}},\ \bibinfo {pages} {403--427} (\bibinfo
  {year} {2021})}\BibitemShut {NoStop}%
\bibitem [{\citenamefont {Ams\"uss}\ \emph {et~al.}(2011)\citenamefont
  {Ams\"uss}, \citenamefont {Koller}, \citenamefont {N\"obauer}, \citenamefont
  {Putz}, \citenamefont {Rotter}, \citenamefont {Sandner}, \citenamefont
  {Schneider}, \citenamefont {Schramb\"ock}, \citenamefont {Steinhauser},
  \citenamefont {Ritsch}, \citenamefont {Schmiedmayer},\ and\ \citenamefont
  {Majer}}]{NVcqed2011}%
  \BibitemOpen
  \bibfield  {author} {\bibinfo {author} {\bibfnamefont {R.}~\bibnamefont
  {Ams\"uss}}, \bibinfo {author} {\bibfnamefont {C.}~\bibnamefont {Koller}},
  \bibinfo {author} {\bibfnamefont {T.}~\bibnamefont {N\"obauer}}, \bibinfo
  {author} {\bibfnamefont {S.}~\bibnamefont {Putz}}, \bibinfo {author}
  {\bibfnamefont {S.}~\bibnamefont {Rotter}}, \bibinfo {author} {\bibfnamefont
  {K.}~\bibnamefont {Sandner}}, \bibinfo {author} {\bibfnamefont
  {S.}~\bibnamefont {Schneider}}, \bibinfo {author} {\bibfnamefont
  {M.}~\bibnamefont {Schramb\"ock}}, \bibinfo {author} {\bibfnamefont
  {G.}~\bibnamefont {Steinhauser}}, \bibinfo {author} {\bibfnamefont
  {H.}~\bibnamefont {Ritsch}}, \bibinfo {author} {\bibfnamefont
  {J.}~\bibnamefont {Schmiedmayer}},\ and\ \bibinfo {author} {\bibfnamefont
  {J.}~\bibnamefont {Majer}},\ }\bibfield  {title} {\enquote {\bibinfo {title}
  {Cavity {QED} with magnetically coupled collective spin states},}\ }\href
  {https://doi.org/10.1103/PhysRevLett.107.060502} {\bibfield  {journal}
  {\bibinfo  {journal} {Phys. Rev. Lett.}\ }\textbf {\bibinfo {volume} {107}},\
  \bibinfo {pages} {060502} (\bibinfo {year} {2011})}\BibitemShut {NoStop}%
\end{thebibliography}%

\end{document}